# Blockchain in Cyberdefence:

# A Technology Review from a Swiss Perspective

Luca Gambazzi, Patrick Schaller, Alain Mermoud, Vincent Lenders

Cyber Defence Campus
armasuisse, Science and Technology, Switzerland

**Abstract:**
*Since the advent of bitcoin in 2008, the concept of a blockchain has widely spread. Besides crypto currencies and trading activities, there is a wide range of potential application areas where blockchains are providing the main building block for secure solutions. From a technical point of view, a blockchain involves a set of cryptographic primitives to provide a data structure with security and trust properties. However, a blockchain is not a golden bullet. It may be well suited for some problems, but often an inappropriate data structure for many applications. In this paper, we review the high-level concept of a blockchain and present possible applications in the military field. Our review is targeted to readers with little prior domain knowledge as a support to decide where it makes sense to use a blockchain and where a blockchain might not be the right tool at hand.*

3 March 2021

**Corresponding author:** luca.gambazzi@ar.admin.ch

# 1 What is a Blockchain?

In 2008, in his seminal work on Bitcoin [1] Satoshi Nakamoto introduced a data structure ("a chain of blocks") as well as a consensus mechanism that enables a set of entities to maintain the general ledger of a currency in a distributed manner. The construction provides security guarantees as long as more than half of the entities participating are honest.

Parts of the difficulty and confusion when talking about "blockchains" stems from the fact that there is no precise definition of what a "blockchain" is. Some consider the whole ecosystem, including all its components, such as the consensus mechanism, the execution environment for a scripting language running on the participating nodes, etc. as "the blockchain", others restrict the focus on the underlying data structure that consists of blocks that contain the data and build a chain. Blocks are tied-up using cryptographic primitives in such a way that it is impossible to modify the blocks' content or to rearrange the blocks, thus resulting in an immutable chain (Figure 1).

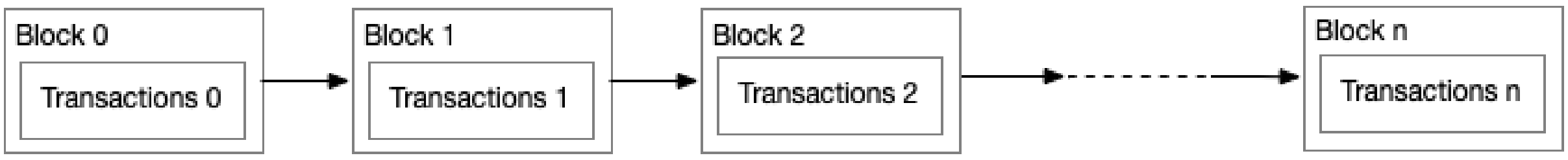

*Figure 1 Immutable Chain of Blocks.*

**First application of the blockchain: Bitcoin**

Nakamato has implemented the first blockchain in 2009 as part of the bitcoin system, where the blockchain represents the core component of the cryptocurrency. In terms of a distributed cryptocurrency such as bitcoin, where the blockchain (here the data structure) is supposed to provide a ledger archiving every single transaction of bitcoins, the requirements are rather obvious:

- **Consensus** on the extension of the blockchain among honest players: every new block added to the chain of blocks must fulfill a set of consistency rules, e.g., transactions contained in a new block have to be valid transactions.
- **Public verifiability**: given data within the blockchain, it is easily verifiable, that the data is indeed part of the blockchain.
- **Immutability** of blockchain data: Once players have reached consensus about the extension of the chain of blocks and once a block was appended to the chain, it should be impossible to change the contents of the block or to rearrange the sequence of the blocks in the chain.

These are probably the most striking properties introduced by the blockchain construction. Of course, there are many other properties to be met, such as efficiency, authenticity, non-repudiation; however, the other properties are more standard, in the sense that there exist other well-established database solutions.

# 2 How a Blockchain operates

## 2.1 Architecture of a blockchain

In the following, we provide more background on the blockchain architecture from Hardware to Application layer (Figure 2) as suggested in [2].

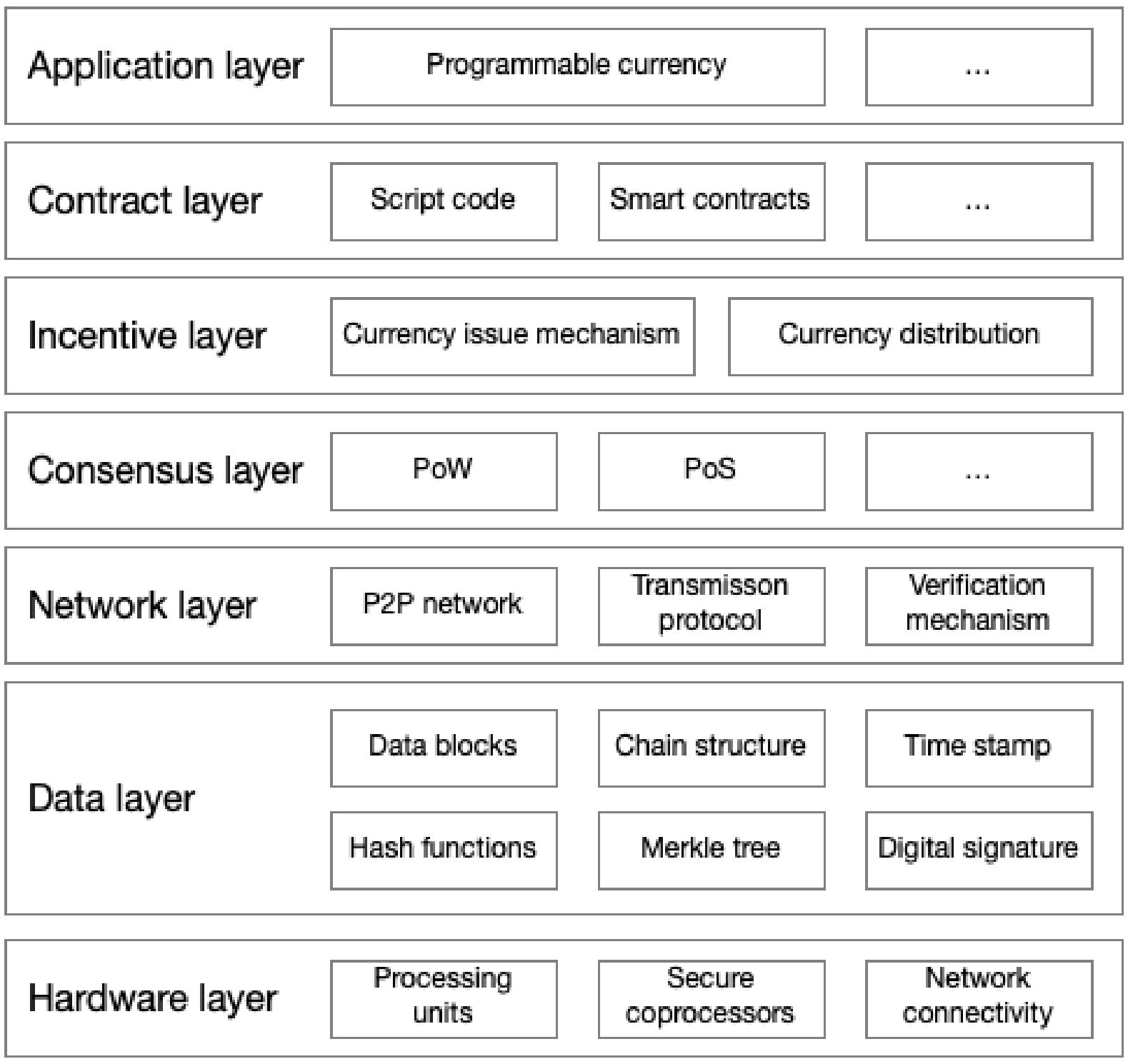

*Figure 2 Blockchain architecture.*

**Hardware layer**

The hardware layer specifies the characteristics necessary for a node to perform blockchain operations efficiently and safely. The requirements to be considered are mainly related to the performances required by the consensus layer, the execution security (e.g., smart contracts, but also the secure use of all the private cryptographic elements) and the connectivity, including the performance requirements of the network layer.

**Data layer**

The **data layer** specifies a data structure that holds the blockchain information. It defines how a valid element of a blockchain (i.e., a "block") looks like, the data structure in a block, and a set of conditions that makes a block a valid element of the blockchain (Figure 3).

Consider for example an element (a block) of the bitcoin blockchain[1]. It contains a set of parameters (e.g., the reference to previous block), as well as the transactions validated in this block (2'580 in the referenced example).

---

1 https://www.blockchain.com/btc/block/00000000000000000000301fcfeb141088a93b77dc0d52571a1185b425256ae2fb

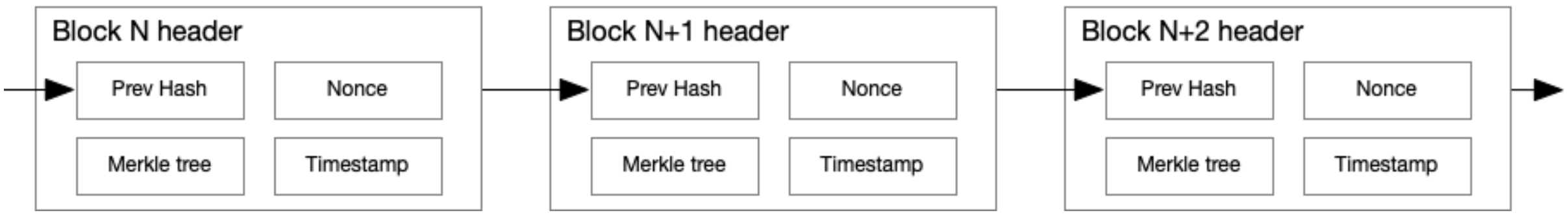

*Figure 3 A simplified view of a blockchain.*

The main characteristic of the data structure used in blockchains is that each block has a unique predecessor and a unique successor, thus building a chain of blocks. The linkage of the data-blocks is implemented using cryptographic tools, such as hash functions and digital signatures, that guarantee that the order of the blocks in the chain is preserved and that the content of the blocks cannot be modified once a block is part of the chain.

### Network layer

The Network Layer defines how elements (nodes) of the blockchain ecosystem communicate and what kind of information they exchange. In the case of the Bitcoin or Ethereum blockchain, the network is a peer-to-peer network[2]. As of October 2020, the Bitcoin network consisted on average of 10'500 nodes reachable from the Internet[3] (Figure 4), the Ethereum network of 8300[4] nodes. A large number of nodes are not accessible directly from the Internet, either because the nodes are elements of private networks (NAT), or because they are part of the Tor anonymity network[5]. In this regard, a work was presented at the Internet Measurement Conference [3].

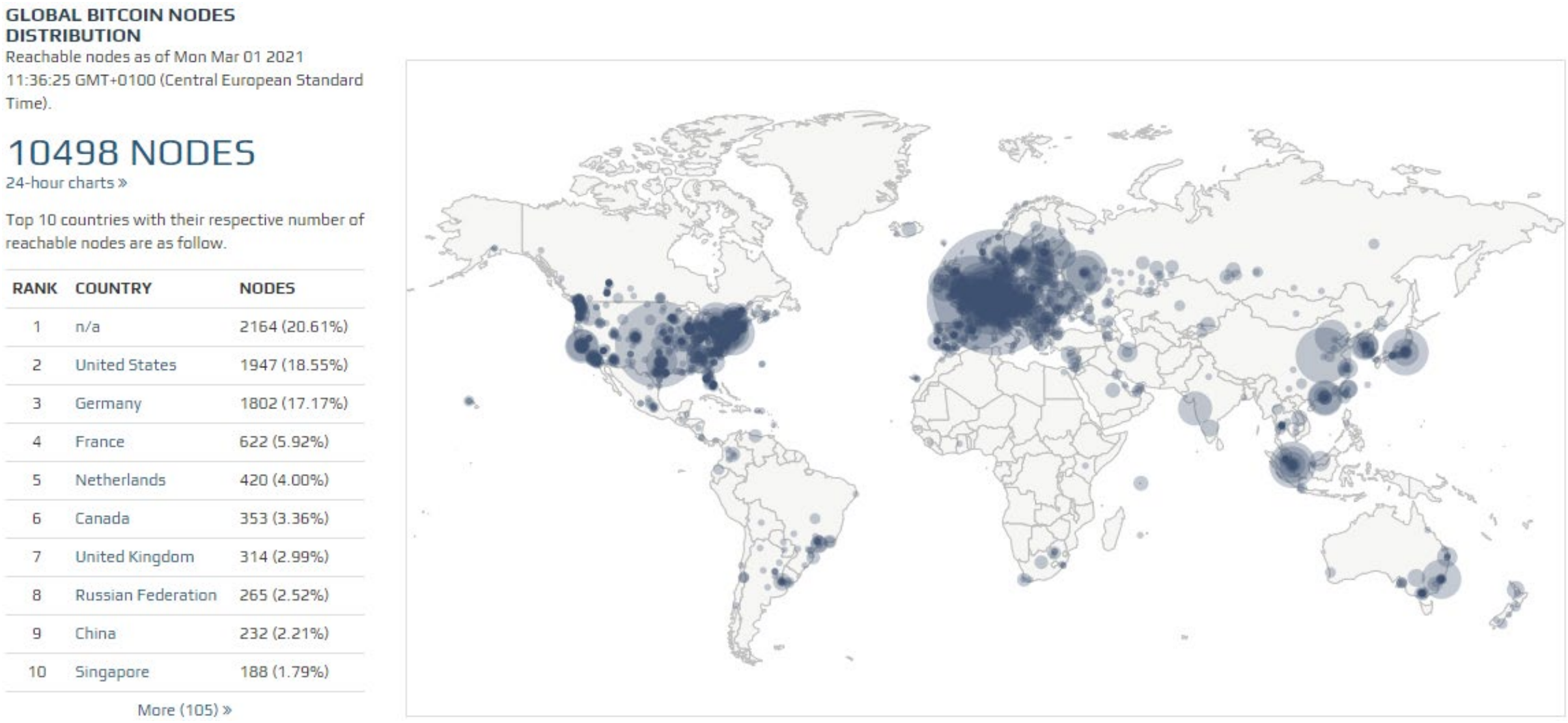

*Figure 4 Reachable bitcoin nodes around the world. Source: bitnodes.io.*

### Consensus

The previous paragraphs have described properties of the underlying data structure and the necessity of a network to build a blockchain ecosystem. Given that we have defined an underlying data structure and network protocols that enable nodes to exchange information, an important follow-up question is how one could achieve agreement about the state and the correctness of the content of the blockchain among the participating nodes especially in the presence of potentially malicious nodes. This problem is trivially solved, if there is a trusted and accepted authority that checks the correctness of the content and returns the current state of the data upon request. One of the strongest properties of the Bitcoin blockchain is the fact that it allows to achieve agreement among honest nodes and correctness of the data in the blockchain in the presence of potentially malicious nodes in a decentralized setting. The keyword here is consensus.

[2] https://bitcoin.org/en/p2p-network-guide
[3] https://bitnodes.io/
[4] https://www.ethernodes.org/
[5] https://www.torproject.org/

"Classic" consensus protocols achieve consensus among honest agents in the presence of malicious nodes, as long as the fraction of malicious nodes does not exceed a certain threshold. A famous result in computer science[6] shows that consensus can be achieved among honest agents if the fraction of malicious agents does not exceed 1/3.

One of the central building blocks of the bitcoin ecosystem and one of the main contributions of Nakamoto [1] is the introduction of a new consensus mechanism, the so-called "Nakamoto consensus". This type of consensus allows the blockchain ecosystem to provide guarantees about the state and the correctness of the blockchain in a decentralized setting, i.e., in a peer-to-peer setting, where there is no central, trusted authority.

"Permissionless blockchains" (decentralized blockchains, presented in detail below) such as Bitcoin or Ethereum rely on this type of consensus mechanisms, the two most well-known are:

- **Proof of work (PoW)**: Extending the blockchain with a new valid block requires a predefined amount of computational work to be completed. The work is designed in way that is hard to be done, but can be easily verified. Similar to solving a puzzle, where assembling the puzzle is hard, checking the correctness of an assembled puzzle is easy. Completing the work and extending the blockchain with a valid block is rewarded (mining). Creating valid blocks for the blockchain thus requires computational power. Given the chain structure of the blockchain, changing a block within the blockchain requires changing all subsequent blocks and thus requires even more work. The simple rule "the longest blockchain is the valid blockchain" creates a race between malicious nodes and honest nodes. It can be argued that as long as more than half of the computational power is controlled by honest players, the correct blockchain grows faster than any maliciously modified version of the blockchain. Nevertheless, it is worth remembering that selfish mining attacks could reduce the tolerable number of malicious nodes in Bitcoin network [4], [5].
- **Proof of stake (PoS)**: Here the node to extend the blockchain with a new block is selected from the set of participating nodes through a combination of earned credits and randomness. The intention is to design or tune the selection criteria in a way such that correct behavior is rewarded, malicious behavior is penalized. Finally, malicious behavior should not be profitable. Note that most of the real-world and open blockchains currently rely on PoW.

### Incentive layer

To guarantee persistence and liveness of a network, it is crucial to attract as many nodes as possible willing to participate in the consensus protocol. To achieve this goal, there have to be incentives for nodes to participate in the blockchain extension process.

In PoW and PoS mechanisms mentioned above, nodes that extend the blockchain and thereby contribute to correct functioning of the system, receive a reward for their contribution.

[6] https://en.wikipedia.org/wiki/Byzantine_fault

In the Bitcoin ecosystem this process is called "mining", since with the creation of a new block the creator is rewarded with a predefined number of newly created (mined) Bitcoins. The reward rule was defined by the Bitcoin creator and is a central rule in the system. The rule cannot be changed without agreement of the entire Bitcoin network. The block reward started at 50 BTC in block #1 and is defined to halve every 210,000 blocks. Thus, every creation of a valid block up to block #210,000 has been reward 50 BTC, while block 210,001 has been rewarded with 25 BTC. This is the only way Bitcoins are created. Consequently, the total number of Bitcoin cannot exceed the limit of 21 million BTC.

Besides the "mining" of fresh coins, creators of valid blocks are paid transaction fees, such that in case all coins have been mined, there is still an incentive to participate and contribute to the ecosystem.

**Contract layer**

Transactions of Bitcoins consist of one or more inputs and outputs. An input denotes a previous transaction that transfers Bitcoins to the payer (of the current transaction), an output assigns Bitcoins to the payee of the transaction. In the Bitcoin ecosystem inputs/outputs of transactions are defined as function associated with the transaction, so called scripts.

For example, a transaction could have a timestamp associated with it, that allows the transaction to be pending and replaceable until an agreed-upon future time, specified either as a block index or as a timestamp.

Bitcoin has a basic set of instructions that can be used to define constraints/conditions on the execution of the corresponding transaction. These scripting languages enable more sophisticated smart contracts. Smart contracts are programs stored in blocks of a blockchain that are executed if a set of conditions is fulfilled. This enables contractual partners to automate the execution of tasks without a third-party intermediary.

Technically Bitcoin supports smart contracts too, but the scripting language is extremely limited making this type of feature impractical. Other blockchain frameworks contain rich scripting languages: for example, Solidity on Ethereum. Smart contract technology could therefore speed up business processes, reduce operational errors, and improve cost efficiency.

**Application layer**

The application layer is the point where users interact with the blockchain. In case of the Bitcoin framework, this would for example be a Bitcoin-Wallet that enables a user to transfer money to another user's Bitcoin-Wallet.

## 2.2 Changes in the architecture

The specification and implementation of the layers presented in the previous paragraphs of this chapter define a blockchain system. As we had explained, it is not enough to define the data structure of a block only, but also requires the definition of the consensus mechanism, the way nodes communicate, a language to define contracts, etc. This defines the rules of the game, which have to be followed by the nodes participating in the system. As such, the rules define/specify the elements in the blockchain system, for example, the client-software that allows end users to take part of the system.

As it is the case for all data processing systems, the requirements might change over time or there may be bugs in the implementation that need to be fixed. As a consequence, data structures or code may need to be changed and the system needs to be updated. Especially in the case of decentralized systems (e.g., Bitcoin) this kind of changes/updates may be challenging. Changes/updates that affect the consensus rules of the system are called *forks*. In some sense they introduce "forking points" into the chain of blocks since after the change/update the governing rules of how the blockchain is extended chances.

One differentiates between *soft forks* and *hard forks*. In case of a soft fork, nodes do not require to update to maintain consensus, because blocks of the "new"/forked blockchain follow the old as well as the new consensus rules. However, nodes that did not update might not be able to produce new, valid blocks since the old set of consensus rules may violate the new set of consensus rules. In case of a hard fork, the new set of consensus rules is not compatible with the old set of consensus rules. Consequently, all participating nodes are required to upgrade to the latest version in order to be compatible with the new version of the consensus rules.

## 2.3 Types of blockchains

Considering a blockchain as a distributed ledger [1], there exist two main types of blockchains:

- **Permissionless blockchains**: Examples of this type of blockchain are Bitcoin [1] and Etherum [6]. The system is open and decentralized and allows any peer to join the network as a reader or writer. There is no membership management, which would ban malicious readers or writers. Typically, the consensus mechanism prevents malicious behavior and guarantees security requirements up to a certain threshold of malicious peers.
- **Permissioned blockchains**: In this type of blockchains only an authorized set of writers and readers is admitted participating in the system. Examples of this type of blockchains is Hyperledger Fabric[7]. Here a consortium of peers or a central entity assigns rights to peers that want to participate in the network.

  In terms of permissioned blockchains, we further distinguish between public permissioned blockchains and private permissioned blockchains. Whereas private permissioned blockchains restrict access to stored data to a set of peers that hold the necessary access rights, public permissioned blockchains allow every to access the stored data and thereby to verify that the data has been stored according to the consensus rules of the blockchain.

Note at this point, that verifiability of the content of a blockchain may interfere with privacy requirements. However, in many cases privacy is achieved by the use of cryptographic techniques to prevent leakage of private data when public verifiability is a requirement.

[7] https://www.hyperledger.org/projects/fabric

# 3 Blockchain market analysis

Blockchain technology can be integrated into multiple application areas. The primary use of blockchains today is as a distributed ledger for cryptocurrencies, most notably bitcoin. There are a few operational products maturing from proof of concepts [7] [8]. Here follows a market analysis with the aim of measuring the presence of this technology on the innovation sector and more specifically in Switzerland.

## 3.1 Blockchain's emergence

Various studies estimate the future potential of blockchains and distributed ledgers as high to very high. In a 2018 study, the World Economic Forum (WEF)[8] calculated world-wide efficiency gains of around 1‘000 billion USD for trade finance alone. Other areas of application include logistics (Maersk plans to optimize its container logistics), retail (IBM and Walmart are developing a solution for food safety), insurance (B3i is developing a smart contract solution for insurance contracts), energy (Axpo is developing a solution for peer-to-peer energy markets), transportation (Novotrans stores inventory level data for railway repairs) or public administration (the Netherlands are developing a border control system for passenger data).

However, businesses have been thus far reluctant to place blockchain at the core of the business structure[9]. Despite the hype, blockchain is still an immature technology, with a market that is still nascent and a clear recipe for success has not yet emerged. Unstructured experimentation of blockchain solutions, without strategic evaluation of the value at stake or the feasibility of capturing, means that many companies will not likely see a return on their investments.

A study of the Internet of blockchain foundation[10] summarizes several applications using blockchain. It should be noted that, as will be discussed in the following chapters, the use of the blockchain is rarely essential: in most of these applications, the blockchain is used as a distributed logbook.

Such applications can also be transposed to Switzerland. Further examples in this country include Modum (pharmaceutical supply chain), Swiss Prime Site (property management and rentals) and UBS (Utility Settlement Coin, trade finance, etc.). Next to improving efficiency, blockchains and distributed ledgers also open up numerous new fields of business, including new services (e.g., digital identity), software development (e.g. new web services or so-called “distributed apps” or “dApps”) and specialist services (e.g. legal). Their successful implementation hinges on at least three critical factors: the availability of talents and their training at institutions of higher education, a well a functioning ecosystem of institutions of higher education, established players and startups (with good access to venture capital), as well as a flexible regulatory and legal framework. While Switzerland has a well-functioning ecosystem, it needs to catch up as regards the education of talents and access to venture capital.[11]

[8] https://www.weforum.org/agenda/archive/blockchain/
[9] https://www.ft.com/content/c905b6fc-4dd2-3170-9d2a-c79cdbb24f16
[10] https://medium.com/@essentia1/50-examples-of-how-blockchains-are-taking-over-the-world-4276bf488a4b
[11] https://www.satw.ch/en/cybersecurity/technology-outlook-2019/

Some probably overoptimistic financial forecasts on statista.com[12] suggest that global blockchain technology revenues will experience massive growth in the coming years, with the market expected to climb to over 39 billion U.S. dollars in size by 2025. The financial sector has been one of the quickest to invest in blockchain, with over 60 percent of the technology's market value concentrated in this field.

The United States, Russia, China, and most of the G20 countries have devoted resources to blockchain solutions. The United Arab Emirates, led by its tech-hub in Dubai, aims to be the world's first blockchain powered government; and the Australia National Blockchain aims to move the nation towards blockchain immersion. Many industry leaders have pooled into consortia — technology-specific, such as R3 and the Ethereum Enterprise Alliance; or business specific, such as the Hyperledger, Bankchain, TradeLens, and MediLedger. The goal for such co-opetition is to bring standards that lifts all boats, such as a Blockchain in Transport Alliance for supply chain management, or a Blockchain Law Consortium for the legal industry.

## 3.2 Blockchain in academia

Clarivate Analytics' Web Of Science[13] provides access to multiple databases including comprehensive citation data for many different academic disciplines. Data retrieved from Web of Science indicate a clear interest and important volume of investment from Chinese institutions. While the number of academic publications from North American, European and Chinese institutions is comparable (~ 1'500 publication / year), it is impressive observing the massively higher amount of Chinese investment: almost 10 times larger than European or North American countries. This could be interpreted in two ways: either as a brute force approach to generate scientific success by funding local institutions disproportionately, or as a strong signal to the world about the means available for Chinese academic development and innovation.

In addition to this information, a quantitative analysis based on title and abstract of the publications on arXiv[14] allows comparing the number of publications covering blockchain or blockchain and cyber (Figure 5).
As of 2018, a clear blockchain-hype is visible, and the first publications related to cybersecurity of blockchain appear. This can be explained by the fact that the security concerns of an emerging technology often appear in a second phase of development, or simply that other application domains, such as cybersecurity or cyberdefence are not correlated with blockchain interest.

[12] https://www.statista.com/statistics/647231/worldwide-blockchain-technology-market-size/
[13] https://clarivate.com/webofsciencegroup/solutions/web-of-science/
[14] https://arxiv.org/

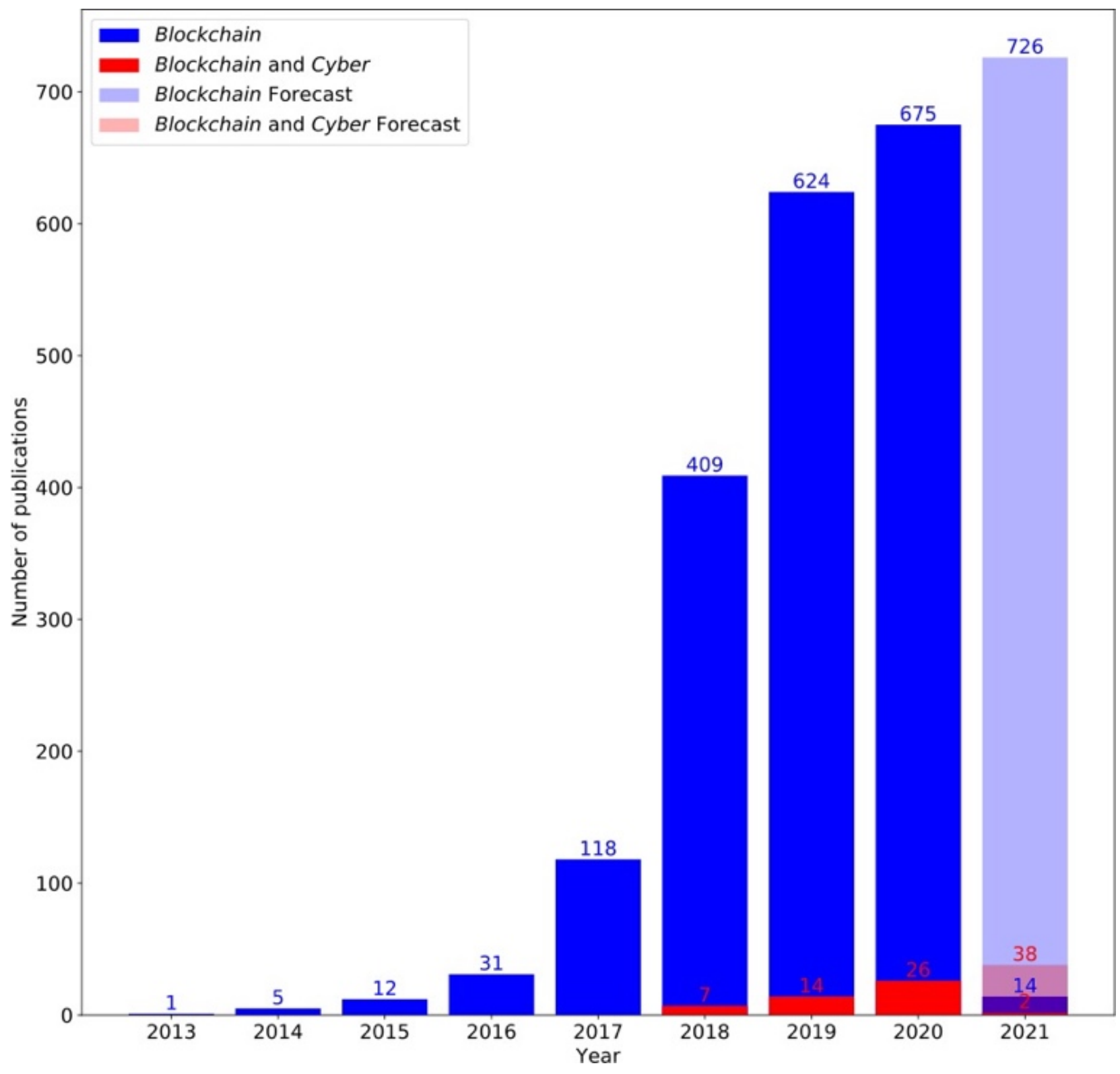

*Figure 5 arXiv publications containing the keywords "Blockchain" (in blue) and "Blockchain AND Cyber*" (in red). Source: S. Gillard, T. Maillart and D. Percia David.*

## 3.3 Blockchain in Switzerland

It is possible to observe that to date, blockchain innovation in Switzerland primarily originates from academia and industry, which in turn are stimulated by the financial sector. The defence sector, on the other hand, does not seem particularly active. Swiss institutions of higher education operate various research centres on the topic of "blockchain"; the EPFL and the ETH Zurich as well as the Universities of Basel, Lucerne and Zurich are very active in this field. The Swiss Confederation's press release "Federal Council wants to further improve framework conditions for blockchain/DLT", published on 14 December 2018, also triggered many social media posts.[15] During its session on 19 June 2020, the Federal Council took note of the report on the need to amend tax law with regard to blockchain. The report concluded that no special legislative amendments to tax law are necessary.[16] The interest in blockchain in Switzerland is presented in the following indicators: analysis of jobs vacancies in Switzerland (Figure 6) and by geographical region (Figure 7)[17].

[15] https://www.admin.ch/gov/en/start/documentation/media-releases.msg-id-73398.html
[16] https://www.efd.admin.ch/efd/en/home/dokumentation/nsb-news_list.msg-id-79513.html
[17] Figure 6 and 7 are based on the automated Technology and Market Monitoring (TMM) system developed by armasuisse S+T. The data is collected by a web crawler that searches through publicly available sources, such as commercial registers, company websites and social media channels. The data is searched at regular intervals and updated monthly by TMM. Thanks to TMM, companies can be located along with relevant information such as the products, services and technologies that they offer.

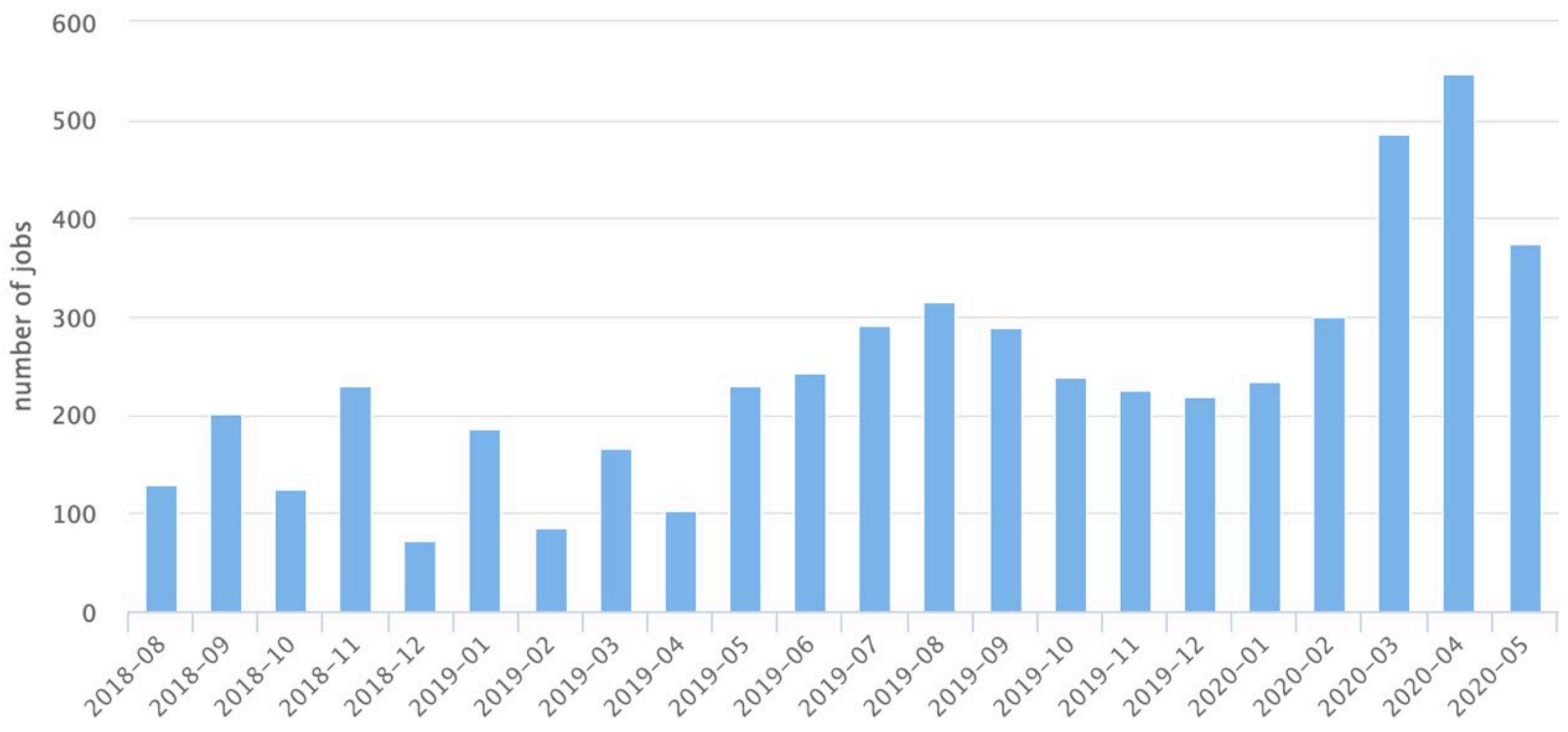

*Figure 6 Number of blockchain jobs opening in Switzerland.*
*Source: TMM armasuisse S+T*

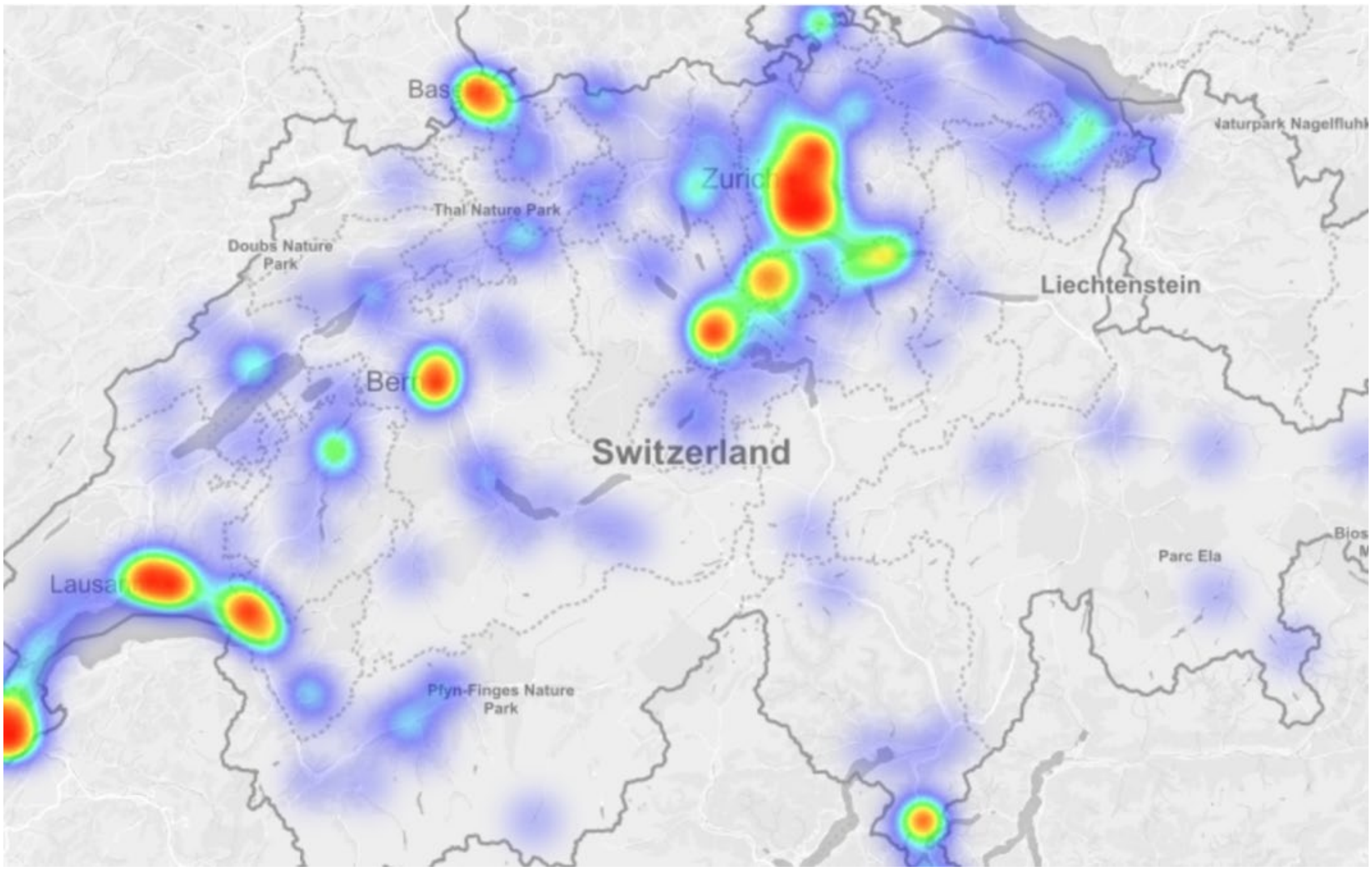

*Figure 7 Blockchain activities based on companies' websites in Switzerland.*
*Source: TMM armasuisse S+T*

## 3.4 A brief history of the Crypto Valley

Switzerland has become a center for new business ideas in the field of blockchain and distributed ledger technology (DLT). In particular, originally based in the canton of Zug, Crypto Valley has established a worldwide reputation as a hub for global growth, resulting in a high density of blockchain and DLT companies throughout Switzerland. The country is known worldwide for its privacy-conscious legislation, world-class talent and its openness. Like blockchain, Switzerland is also organized in a decentralized way, which has a positive effect on understanding this new technology. The state government's open and proactive attitude has led to favorable conditions for blockchain companies, which has created an ecosystem that produces world premieres: in 2016, Zug became the first city in the world to accept Bitcoin payments for tax purposes; in 2017, Crypto Valley announced the introduction of a decentralized Ethereum-based digital ID system; and in 2018, the fintech company Amun launched the world's first crypto index product on the SIX Swiss Exchange.

## 3.5 Blockchain and patents

In this market analysis, we have deliberately omitted patent data analysis. Well aware that in order to identify new competitive products and processes it is necessary to have access to detailed information on technological innovations, but analysis of software patents is often misleading. While for other products analysis on filed patents could be an effective method to obtain such information, for example by indicating the types of products and processes that companies are planning to introduce to the market; when it comes to software, quantitative patent analysis distorts the reality. In fact, the vast majority of blockchain-based solutions are based on open-source technologies, and the legislation governing the patentability of software is extremely heterogeneous globally [9]. In this sense, we observe the largest number of patents in countries (such as China[18]) where patenting software is simple.

[18] We observe that China fills most of the blockchain-related patents (49%). The US is the second chosen jurisdiction for filing blockchain patents (19%); a 12% of patents documents are worldwide applications that have gone via Patent Cooperation Treaty (PCT) through the Word Intellectual Property Office (WIPO). Korea and Japan are also leading countries for blockchain patent filings; other key regions where protection is sought are Australia, Canada, Taiwan, India and Singapore. Great Britain is the first European country for blockchain patents, followed by Germany.

# 4 Can a blockchain solve your problem?

In the previous chapters, we provided a brief description of the components and properties of a blockchain and a market analysis as well. Many of the component observed are not exclusive to the blockchain technology: distributed databases exist in many forms, cryptographically signed chain of events too, etc. Here we present three steps that allow you to decide whether a blockchain is the right technology to solve your problem or not.

## 4.1 Storing information in an untrusted environment

In [10] the authors present a decision graph that allows you to decide whether a Blockchain is the right tool to solve a given problem. They consider the three main types of blockchain (permissionless, public permissioned and private permissioned) and a guide for decision-making (Figure 8).

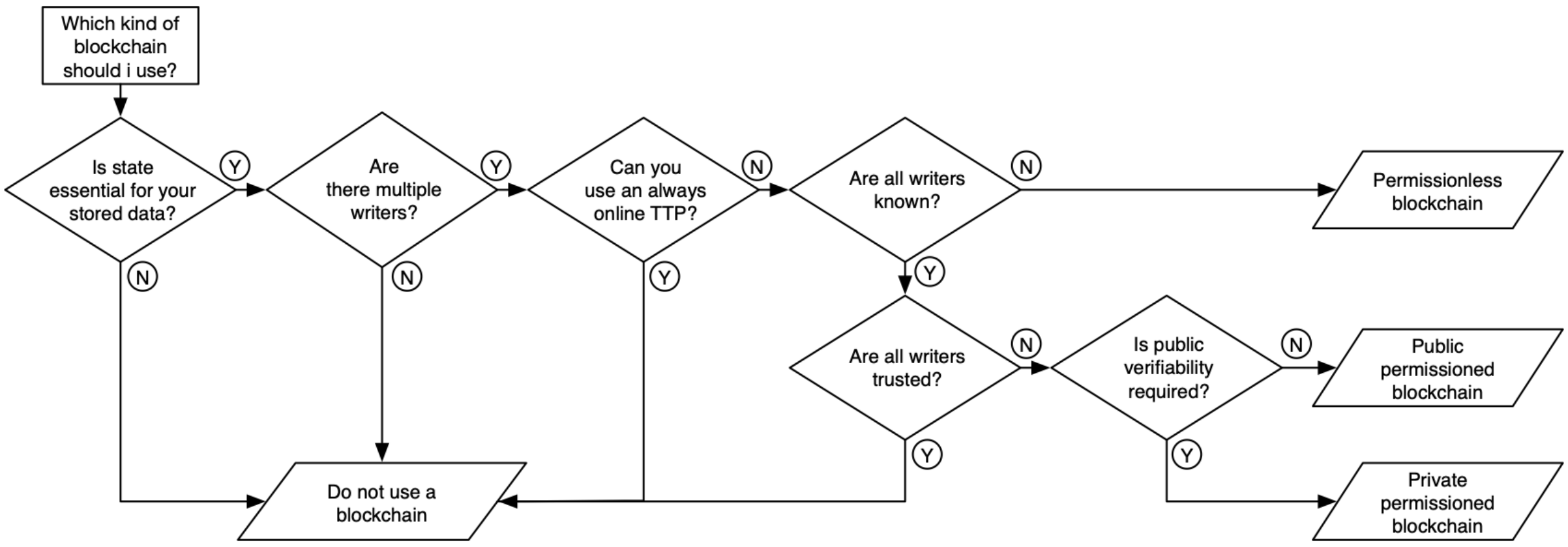

*Figure 8 Decision graph: choosing the right Blockchain type in an untrusted environment.*

**Necessity to store state changes: is state essential for your stored data?**

A blockchain, as a form of a database, stores information in sequential order and ensure the integrity of the data and of its order. In this context, "is state essential" indicates the need to store information in a given sequential order and to protect its integrity. This must be an essential requirement in order for the blockchain data structure to make sense.

**Multiple writers: are there multiple writers?**

Although straightforward, it is worth remembering that if only a single entity is responsible for writing data, the integrity and authenticity of the data can be guaranteed without the use of a blockchain.

**Trusted third party: can you use an always-online TTP?**

If there is a trusted third party (TTP), this can verify the state changes and guarantee transitions correctness. The TTP could be used as a trusted writer in case the TTP fulfills the availability requirements. In this case, from a purely technical point of view, a database would be a better and more performing solution.

In case the TTP does not meet the availability requirements, the TTP, acting – for example – as a Certificate Authority, can establish a group of trusted writers. The TTP in that sense ensures that the security requirements are met.

**Known writers: are all writers known?**

If participants do not know each other, and there is no agreement on a common trusted third party, the solution – as in the case of cryptocurrencies – is an open blockchain.

**Trusted writers: are all writers trusted?**

If all the writers are trusted there is obviously no need for a system that guarantees integrity and sequential order of entries, since by definition writers act according to the rules.

On the other hand, if we consider the case that writers could be malicious and try to compromise the integrity of the stored data and the correctness of each transaction, a blockchain could be the ideal solution.

**Public verifiability: is public verifiability required?**

Finally, it is possible to allow access to the blockchain, without the possibility of changing the chain status, to allow third parties to verify the state of the data saved, and consequently their correctness at all times.

## 4.2 Interaction with the physical world

Depending on the type of blockchain chosen with the help of the decision graph previously presented, we find ourselves in a situation where writers are either unknown, not trusted and therefore potentially malicious. Hence, we provide an additional set of observations with the aim of making the reader aware of the limits of the blockchain when interaction with the physical world becomes necessary (Figure 9).

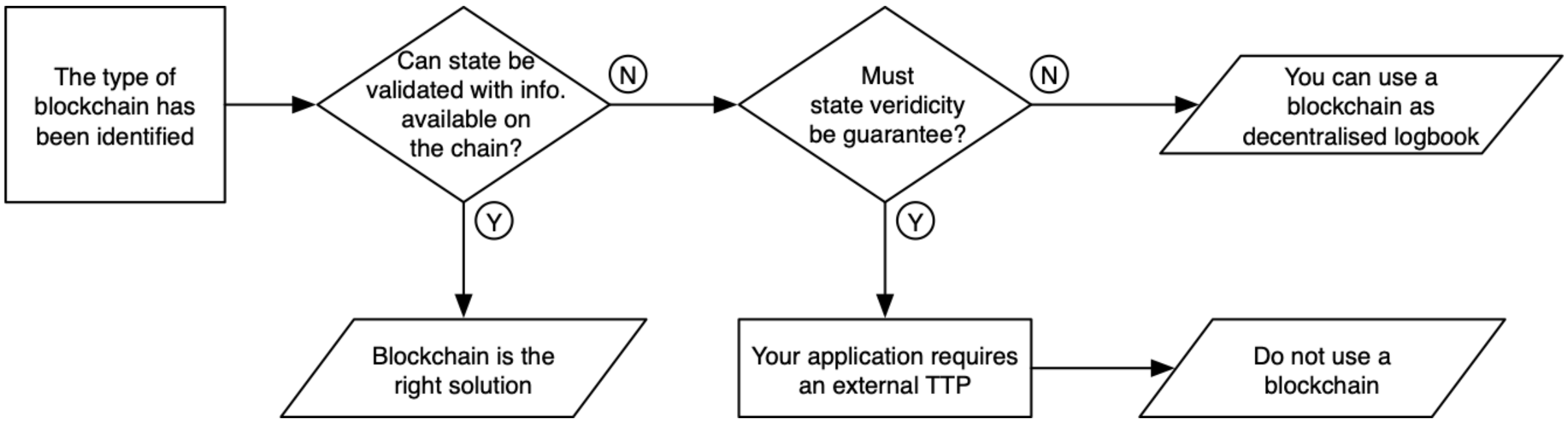

*Figure 9 Using a blockchain when interacting with the physical world.*

**Can state be validated with information available on the chain?**

It is necessary to be aware of one additional, fundamental aspect before choosing a blockchain as a technical solution: a blockchain can guarantee that an entry in the ledger reflects the corresponding state only if data and its state changes can be validated with information available on the chain itself. This is the typical scenario in financial applications, where for instance the creation of crypto currencies are the product of the mining process and therefore no interaction with the physical world is ever required.

**Consensus and integrity: must state veridicity be guarantee?**

Consequently, if someone wants to capitalize on the characteristics of a blockchain as a whole, he has to find a use case where both data and its state changes can be validated with information available on the chain and they do not depend from external sources.

If we consider the two main security properties provided by a blockchain as

1. consensus about the current state of the blockchain
2. guarantee that the blockchain cannot be changed afterwards

there is an interesting requirement of cryptocurrencies: in the cryptocurrency case, where the blockchain provides some sort of a ledger, the user does not care about the honesty of the writers. If the ledger contains a transaction according to the user expectations, he has by property 1 the guarantee that everybody agrees that he owns now the money and by property 2 that there is no way the money could be taken from his wallet. In that sense the main value is that it is in the blockchain and that nobody can change it.

Following the decision graph presented above, as soon as this kind of "guarantee of veridicy" or "trust" exists, there is the question if the blockchain is then still the right tool, since then one could most probably also trust that the writer does not change it afterwards.

**Decentralized logbook**

On the other hand, in many non-cryptocurrency applications, the main role of the blockchain is to provide some kind of a verifiable decentralized logbook where information originated in the physical world are stored on the chain.

Typically, the content of a block is reserved to transactions and smart contracts, but this is not mandatory. It is possible to use a blockchain as distributed storage by adding arbitrary data to each transaction.

Although it is not possible for a blockchain to provide any guarantee that this arbitrary data in the ledger reflects a corresponding state of the reality, this immutable, distributed and secure data model makes it very attractive to add arbitrary information form physical world on a blockchain.

In contrast to the cryptocurrency case, where the main value is the entry in the blockchain itself, in these scenarios, the main value lies in the fact that the content in the blockchain reflects the corresponding state in the physical world. Thus, it requires trust into the one who reported the physical value written in the blockchain.

In case of a violation of the rules in the physical world, the blockchain allows detecting and tracing previous operations, but does not prevent an incorrect write to happen.

### 4.3 Starting a new blockchain or relaying to an existing one

By separating out the architecture of the Blockchain into multiple layers, we could better study the various properties that we want the Blockchain to enjoy and where they need to be implemented. Lastly, before choosing the right solution for a blockchain other aspects have to be taken in account:

- **Security**: no party should be able to control a majority of some scarce resource (typically computing power), and therefore to convince nodes that an alternate version of the ledger is the valid one,
- **Liveness**: nodes can add new blocks to the ledger with acceptable latency,
- **Stability**: nodes in the network should not alter their opinion of the consensus ledger (except in very rare cases),
- **Correctness**: only blocks that represent valid transactions (i.e. they conform to a specification of how new blocks may relate to previous blocks) may be added to the ledger.

Because of all these characteristics, when blockchain fits your requirements, you will have to choose carefully if you want to create your own new blockchain or to use an existing live, solid and stable chain.

# 5 Blockchain in defence: Military application of blockchain

Based on the elements and decision-making processes presented above, it is now possible to assess whether the blockchain is the right solution for an application. In this chapter, we carry out this type of analysis systematically.

As presented in the previous chapter, using an open or permissioned blockchain is justifiable when multiple mutually mistrusting entities want to interact in order to change the state of a system, and are not willing to agree on an online trusted third party.

Among the most cited applications in the military press, we present three of them covering supply chain, detached labels and messaging applications. Even for these specific examples, the three applications have not reached a sufficient stage of maturity into the military domain today.

Before analyzing individual cases, it is important to revisit two fundamental concepts, as stated in [10] and [11]; using a blockchain in a particular application scenario make sense when:

1) the scenario requires multiple mutually mistrusting entities to interact and record or change some state of a system, and
2) in respect to use of a common online Trusted Third Party (TTP): the entities are not able to agree on such TTP or the implementation of such a common TTP is not possible.

It is obvious that the first condition is not the case in most of the national and alliance peacetime operations, as usually there are some pre-existing trust arrangements between the parties involved in the interactions. Nevertheless, situation is different when it comes to federated mission operations.

## 5.1 Supply chain for logistics and procurement

**Scenario:** The modern military logistics and supply chain brings together hundreds of different military and private sector components [12]. With so many participants, there are numerous points of friction that introduce numerous failure points, unnecessary costs, and result in inaccuracies and misrepresentation. By providing a single source of truth and supporting intelligent automation, blockchain is a technology candidate to address these challenges and allows keeping track the origins and history of transactions in various commodities.

Moreover, integrating blockchain within each step of an operation to secure and share data throughout the manufacturing process, including design, prototyping, testing, and production; deployment blockchain may offer the defense procurement a solution[19].

Using a public infrastructure such a blockchain could ease integration with non-military partners, as well as facilitates civil-military cooperation without depending of a common trusted authority (or cross-signed PKI certificates sharing multiple authorities).

**Choosing a blockchain:** in a logistic and procurement environment, storing state (e.g. equipment inventory) is critical; moreover, a supply chain can be complex and requires multiple writers. If a trusted third party is agreed among all the writers, a blockchain is not necessary; on the other hand, if it is not possible to agree on a TTP a blockchain-based solution might make sense. In case of a supply chain, depending on the confidentiality of the saved information, you can choose the blockchain model.

[19] https://www.army.mil/article/227943/blockchain_for_military_logistics

**Interaction with the physical world:** in this particular case, the correctness of the information entered in the blockchain is crucial; it would be a failure if the information stored did not correspond to reality. To solve this problem, we could then delegate the responsibility to inventory the assets to an external body or use certified sensors, which ensure - for example - the inventorying of the assets. In the first case, this means we would agree on a TTP responsible for data entry; in the second case we would accept data entry only by certified sensors (trusted writers): in both cases, we would find ourselves in a situation where the blockchain is no longer needed.

For this type of application, the use of a database (centralized or decentralized) would be more effective and performant.

## 5.2 Detached labels and proof of ownership in a federated environment

**Scenario:** in [11], the author suggests using a distributed ledger for implementing a distributed solution for storing detached labels. The proposed solution aims to add a tool for information security in organization. Hence, ensure traceability and offer a transparent service to verify a metadata collection for classified information (e.g. ownership, classification, expiration).

Such labels include metadata-describing data objects as defined in STANAG 4774. Metadata are neither directly attached to nor stored with the data objects. For such application, it seems to be efficient to write this information in a blockchain as detached metadata instead of depending of a PKI infrastructure.

In particular, authors propose to allow only data originators to assign security marking to a document (proof of ownership). Moreover, the blockchain shall be publicly readable, so that any party requiring access to the data or receiving the data could be able to verify the security requirements for handling and protection of a particular data object. The security classification itself can change over time: for example, due to a de-classification of information after some period of time.

As an additional application in the physical world, the same infrastructure could be used to keep an inventory of the physical copies of a sensitive document, recording new item creation (e.g. reproducing documents), as well as their destruction and confirmation of destruction.

**Choosing a blockchain:** In this specific case, it is essential to save the status of the information, allowing the different data originators to enter information. Moreover, the author assumes that it is not possible to have a TTP always online.

The choice of an open or permissioned blockchain is left to the user, as long as it is possible to verify the information (unrestricted or limited to a restricted group of clients). Regardless of the solution chosen, it will be necessary to assess the impact of an open or public blockchain on the confidentiality of the metadata entered.

Here, it is assumed that only the data originator, who first registered an object on the blockchain, or the users to whom he delegated the responsibility, can update its state. This is interesting, because it increases - without certain guarantees - the feeling of correctness of the information entered.

**Interaction with the physical world:** the main point in this scenario is the proof of ownership and the consequent exclusive responsibility to update the state of an object in the distributed ledger assigned to the first user who claims its ownership.

To avoid with certainty that a malicious writer cannot claim ownership of an asset that does not belong to him, we should delegate the writes to a TTP or trust all writers, in both cases a blockchain would then not be necessary.

Moreover, it is realistic to think that the status associated with the assets described in the blockchain changes regularly over time: what to do if the data originator can no longer change the status of an asset (for example, no longer having access to the credential used to demonstrate its ownership)? Probably you should decide to have a *super partes* TTP capable of reassigning the ownership of an asset or instead to give up updating the status of an asset.

## 5.3 Messaging

**Scenario:** Since 2016 DARPA invested about 1.8 mio. dollars to study public blockchains use cases[20,21]. DARPA asked experts to develop efficient methods to use blockchain technology to support messaging application.

Objective of the mandate is to use the stable and reliable infrastructure of existing blockchain such Bitcoin or Ethereum as a transport layer for messages. In this specific scenario, DARPA does not aim to use blockchain properties as in the case of crypto currencies, but rather seeks to exploit existing blockchain as a method of transporting messages.

**Choosing a blockchain:** Now, aware that for this type of approach, according to the decision graph in chapter 4, an open blockchain is the solution sought (as independent from an authorization system such as a certificate authority) there are two challenges to solve:

- The first one concerns anonymity: it must not be possible to trace and identify who pushed a message to a node participating in a blockchain. It is good to remember the communication between nodes (miners) participating in a blockchain is different for each solution, as is the fee required for each transaction. As an example, the bitcoin communications are not encrypted, so it would be possible, by analyzing the network traffic on a large scale, to trace the source of a message. On the other side, Ethereum communication between nodes is encrypted (while it is not for the discovery of nodes); in this case, it would probably be possible to leave a message without third parties tracing its origin.
- Secondly, it must not be possible for third parties to retrieve the content of the message. Encrypting a message and depositing it on the blockchain is not enough, as the encrypted message would remain potentially available to analysts forever. In the future, it will be possible to decipher it with the help of more powerful computers, or by exploiting cypher vulnerabilities unknown today.

Based on this latest observation, one can envision exploiting the temporary information used by a network of miners (e.g. discovery protocol) to insert obfuscated messages in the transmission of data between nodes. In this case, the use of the blockchain might actually be interesting. Even so, at this very moment, no pragmatic and efficient solution has been released.

[20] https://sociable.co/technology/darpa-explores-blockchain-to-develop-unhackable-code-for-military/
[21] https://media.consensys.net/why-military-blockchain-is-critical-in-the-age-of-cyber-warfare-93bea0be7619

## 5.4 Further blockchain use-cases in Defence

In 2016, the US Department of Homeland Security (DHS) announced a project that would use blockchain as a means of securely storing and transmitting the data it captures. Using the Factom[22] blockchain, data retrieved from security cameras and other sensors are encrypted and stored, using blockchain as a means to guarantee data distribution, integrity and traceability. The project is still ongoing.

Many other proposals for military applications were analyzed in the preparation of this technology review [13] [14], but none of them essentially requires the fundamental properties to be found in a blockchain.

[22] https://www.coindesk.com/factom-blockchain-project-wins-grant-to-protect-us-border-patrol-data

# 6 Conclusion

In this review, we presented the architecture elements composing a blockchain environment, a series of questions guiding the reader to choose, or not to choose a blockchain technology stack.

Moreover, we presented how to select the right blockchain between permissionless, public permissioned and private permissioned. We also brought to light the benefits and limitations of blockchains, particularly with regard to the integration of information from external sources that are not trusted.

A market analysis from a global and regional perspective highlights many initiatives, these are not related to defence but mainly to the financial world. In addition, there is great interest in using the blockchain as a transparency tool where there is a need to publish data, guaranteeing its integrity and immutability without necessarily being linked to a reference authority.

Many experts bet on the future stability and performance of blockchains, allowing generic information storage and distributed traceability. Especially in government applications, the need of public decentralized verifiability of information where the government itself cannot act as a single, perpetual trusted authority, a distributed ledger could provide a solid base for shared democracy data.

Most of the applications depend on existing chains such as Ethereum, because of its stability, liveness and the ability to integrate complex smart contracts. Nevertheless, solutions like the ones proposed in the Hyperledger umbrella project could dramatically increase the performances of a blockchain network, giving the ability to script more complex smart contracts (e.g. Go, JavaScript) and become the reference in the open source community.

Finally, we presented three interesting examples for defense, extremely different from each other: a supply chain scenario for logistics management, a metadata repository with a focus on information ownership, and the exploitation of a blockchain to convey messages.

In all cases the use of the blockchain is possible but not necessary. In cases where a form of "trust" is necessary, the question always comes back if the use of the blockchain is essential or if a central or distributed database can be a more efficient solution.

From a military point of view, despite the interest aroused by this technology, to date there are no blockchain-based products integrated into defense systems. In particular, the need to have no trusted authority and the awareness that information stored on a blockchain cannot be removed in the future, makes this type of solution not suitable for tactical applications where the control of the life cycle of an information is crucial. The focus goes on federated environments where mistrusting entities need to share information or proof some sort of ownership or presence, but advantages are not evident compared to existing information sharing solutions and could have a negative impact due to the overhead introduced (e.g. infrastructure, responsiveness, energy consumption).

We are convinced that information decentralization, distributed databases and a zero-trust architecture are fundamental elements for future interconnected tactical systems; however, with the elements presented in this document, we can say that the blockchain is not necessarily required to address these design goals.

# 7 Acknowledgements

We would like to thank Arthur Gervais for his feedback and suggestions for this technology review. Moreover, we would like to thank Sébastien Gillard, Thomas Maillart and Dimitri Percia David for Figure 5.

# 8 References, bibliography and additional resources

[1]: S.Nakamoto, "Bitcoin: A Peer-to-Peer Electronic Cash System", https://bitcoin.org/bitcoin.pdf, 2008.

[2]: R. Zhang, R. Xue, L. Liu, "Security and Privacy on Blockchain", ACM Computing Surveys, 2019.

[3]: Seoung Kyun Kim et al., "Measuring Ethereum Network Peers", Internet Measurement Conference, 2018.

[4]: T. Locher, S. Obermeier, Y. Pignolet, "When Can a Distributed Ledger Replace a Trusted Third Party?", IEEE Blockchain, 2018.

[5] Ittay Eyal, Emin Gün Sirer, Majority is not Enough: Bitcoin Mining is Vulnerable, International Conference on Financial Cryptography and Data Security, 2013.

[6]: V. Buterin, "Ethereum Whitepaper", https://ethereum.org/en/whitepaper/, 2013.

[7]: M., Krótkiewicz M., Srinilta C. Intelligent Information and Database Systems. ACIIDS 2020. Communications in Computer and Information Science, vol 1178. Springer, Singapore

[8]: Ford, B. Technologizing Democracy or Democratizing Technology? A Layered-Architecture Perspective on Potentials and Challenges, 2020.

[9]: Bhatt P.C., Kumar V., Lu TC., Cho R.LT., Lai K.K. Rise and Rise of Blockchain: A Patent Statistics Approach to Identify the Underlying Technologies. 2020.

[10]: K. Wüst, A. Gervais, "Do you need a Blockchain?", Crypto Valley Conference on Blockchain Technology (CVCBT), 2018.

[11]: K. Wronay, M. Jarosz, "Does NATO need a blockchain?", NATO Communications and Information Agency, Milcom, 2018.

[12]: Hsieh, M., & Ravich, S.. Leveraging Blockchain Technology to Protect the National Security Industrial Base from Supply Chain Attacks. Research memo, Foundation for Defense of Democracies, 2017.

[13]: Sudhan, A., & Nene, M. J. Employability of blockchain technology in defence applications. In 2017 International Conference on Intelligent Sustainable Systems (ICISS) (pp. 630-637). IEEE.

[14]: Tarhini, A., & Chedrawi, C. Blockchain in the security and defense sector, 2019.